\documentclass[11pt,a4paper,english,showpacs,superscriptaddress,prd,aps,preprint]{revtex4-1}

\usepackage{amssymb}
\usepackage{amsmath}
\usepackage{graphicx}
\usepackage{babel}
\usepackage{hyperref}
\usepackage{color}

\begin{document}

\title{Gravitational corrections to the scattering of scalar particles}

\author{A.~C.~Lehum}
\email{andrelehum@ect.ufrn.br}
\affiliation{Instituto de F\'\i sica, Universidade de S\~ao Paulo\\
Caixa Postal 66318, 05315-970, S\~ao Paulo, SP, Brazil}
\affiliation{Escola de Ci\^encias e Tecnologia, Universidade Federal do Rio Grande do Norte\\
Caixa Postal 1524, 59072-970, Natal, Rio Grande do Norte, Brazil}

\begin{abstract}

We evaluated the scattering amplitude of neutral scalar particles at one-loop order in the context of effective field theory of quantum gravity in the presence of a cosmological constant. Our study suggests that quantum gravitational corrections induce an asymptotic freedom behavior to the $\lambda\phi^4$ theory, for a positive cosmological constant. This result hints that a complete theory of quantum gravity can play an important role to avoid the issue of triviality in quantum field theory.    

\end{abstract}

\pacs{04.60.-m,11.10.Hi}


\maketitle

\section{Introduction}\label{intro}

In a field theory the observable (renormalized) charge might be screened by quantum corrections, in such a way that the classical theory which describes interacting fields becomes a quantum theory of noninteracting particles, or trivial, at finite energy scale. This phenomenon, known as quantum triviality, appears in theories that are not asymptotically free. Examples of non-asymptotically free theories are the Quantum Electrodynamics (QED) and the $\lambda\phi^4$ model. In the case of QED, the energy scale in which the theory becomes trivial is known as the location of the Landau pole. Of special interest is the $\lambda\phi^4$ theory, that is closely related to the Higgs sector of the Standard Model (SM) of particle physics and apparently suffers from a similar problem of the Landau pole. In general, the inclusion of other fields can change the asymptotic behavior of the coupling constants, however the question of quantum triviality in Higgs models is still an open problem. 

Even though the Einstein's theory of gravity is nonrenormalizable~\cite{'tHooft:1974bx,PhysRevLett.32.245,Deser:1974cy}, it can be seen as an effective quantum field theory at energies much below the Planck scale ($M_P\sim10^{19}$GeV$/c^2$) ~\cite{Donoghue:1994dn}. From this point of view, it has been argued that gauge constant couplings become asymptotically free when quantum gravitational corrections are taken into account, even if in the absence of gravity they do not present such behavior~\cite{Robinson:2005fj}. The origin of such effect is the raising of quadratic UV divergences induced by quantum gravity corrections. Such proposal have been criticized by several authors suggesting a possible gauge-dependence \cite{Pietrykowski:2006xy} and the existence of ambiguities in the quadratic UV cutoff dependent part of the effective action~\cite{Felipe:2011rs,Felipe:2012vq}. 

Using dimensional regularization, Toms evaluated the gauge invariant Vilkovisky De-Witt effective action of the QED coupled to the Einstein gravity in the presence of a cosmological constant, showing that quantum gravitational corrections should make the electric charge asymptotically free~\cite{Toms:2008dq}, corroborating the Robinson and Wilczek proposal, but in a different (less drastic) manner. The negative contribution to the beta function of the electric charge should be proportional to the cosmological constant, instead a quadratic dependence on the cutoff. In recent years, several articles have been devoted to the discussion of the gravitational corrections to the coupling constants, including scalar models~\cite{Mackay:2009cf,Chang:2012zzo,Pietrykowski:2012nc}, Yukawa~\cite{Zanusso:2009bs} and gauge couplings~\cite{Ebert:2007gf,Tang:2008ah,Toms:2009vd,Toms:2010vy,Toms:2011zza,Narain:2012te}. In common, all of them employ the techniques of effective action calculations.  

The physical motivations in defining the renormalization of the coupling constants (through effective action computations) have also been questioned by the authors of Ref.~\cite{Anber:2010uj,Donoghue:2012zc}. They argued that only a scattering matrix computations can be used to define the effective coupling constant with a physical relevance. In particular, they have shown that it is not possible to define the Yukawa coupling in an universal way. Considering one-loop corrections with one graviton exchanging to the scattering amplitude of massless fermion-fermion and fermion-antifermion, it has been shown that the running of the Yukawa coupling would be process dependent, i.e., what appears to be an asymptotically free coupling constant in one process, turns out to run in the opposite direction in other.

In this work we investigate the role of the gravitational corrections, in the presence of a cosmological constant, to the scattering amplitude of massless scalar particles at one-loop order. Using dimensional regularization in the minimal subtraction scheme to renormalize the one-loop amplitude, we show that a positive cosmological constant $\Lambda$ modifies the running behavior of the $\lambda\phi^4$ coupling constant in the direction of asymptotic freedom. Because of the magnitude of the observed cosmological constant ($\sim 10^{-47}$GeV$^4$), the quantum gravitational correction to the beta function of $\lambda$ is very tiny presenting no phenomenological consequences, but this result hints that a complete theory of quantum gravity can play an important role to avoid the issue of triviality in quantum field theory. 

The article is organized as follows. In Sec. \ref{scalargravity} we present the model and evaluate the propagators. In Sec. \ref{sec2} we compute the scattering amplitude between two neutral scalar particles in gravitational and self-interaction. We also renormalize the amplitude and compute the beta functions of the coupling constants. In Sec. \ref{conc} we present our final remarks and conclusions.                 

\section{Scalar fields coupled to Gravity}\label{scalargravity}

A model which describes interacting scalar fields coupled to the Einstein gravity can be defined by the following action
\begin{eqnarray}\label{eq01}
S=\int{d^4x }\sqrt{-g}&&\Big{\{}-\frac{2}{\kappa^2}(R+2\Lambda)+g^{\mu\nu} \partial_\mu \phi \partial_\nu\phi -\frac{\lambda}{4!}\phi^4+\cdots\Big{\}},
\end{eqnarray}

\noindent where $\Lambda$ is the cosmological constant and $\kappa^2=32\pi G$, with $G$ being the Newtonian gravitational constant. The dots stand for gauge-fixing and Fadeev-Popov terms, as well as high order terms such as $\lambda_1\phi^2\partial^{\mu}\phi\partial_{\mu}\phi$.

We will consider the gravitational field quantized for small fluctuation around flat metric,
\begin{eqnarray}\label{eq02}
g_{\mu\nu}=\eta_{\mu\nu}+\kappa h_{\mu\nu},
\end{eqnarray}

\noindent where $\eta_{\mu\nu}=(1,-1,-1,-1)$ and $h_{\mu\nu}$ is the graviton field. Even though flat space-time is not solution of Einstein's equation in the presence of a cosmological constant, it is enough to consider it since our goal is to study the quantum corrections to the renormalization of the coupling constant $\lambda$~\cite{Toms:2008dq}. Because of this, the scattering amplitude is evaluated off-shell.  

The metric and its inverse must satisfy $g^{\mu\rho}g_{\rho\nu}=\delta^{\mu}_{\nu}$, therefore the inverse metric and the square root of minus determinant of the metric up to order of $\kappa^2$ can be cast as 
\begin{eqnarray}\label{eq02a}
&&g^{\mu\nu}=\eta^{\mu\nu}-\kappa h^{\mu\nu}+\kappa^2 h^{\mu\alpha}{h_{\alpha}}^{\nu}+\mathcal{O}(\kappa^3),\\
&&\sqrt{-g}=1+\frac{1}{2}\kappa h-\frac{1}{4}\kappa^2h_{\alpha\beta}P^{\alpha\beta\mu\nu}h_{\mu\nu}+\mathcal{O}(\kappa^3),
\end{eqnarray}

\noindent where $P^{\alpha\beta\mu\nu}=\dfrac{1}{2}(\eta^{\alpha\mu}\eta^{\beta\nu}+\eta^{\alpha\nu}\eta^{\beta\mu}-\eta^{\alpha\beta}\eta^{\mu\nu})$ and $h=\eta^{\mu\nu}h_{\mu\nu}$.

Let us employ the harmonic gauge-fixing condition, $G_\mu=\partial^\nu h_{\mu\nu}-\dfrac{1}{2}\partial_\mu h$, which leads to the gauge-fixing Lagrangian \begin{eqnarray}\label{eq03} 
\mathcal{L}_{gf}=\eta^{\mu\nu}G_\mu G_\nu -\bar{c}_\mu \left(\frac{\delta G^\mu}{\delta\xi^\nu}\right)c_\nu, 
\end{eqnarray}

\noindent where $\bar{c}_\mu$ and ${c}_\nu$ are the ghost fields.

Since we are interested in study the scattering of the scalar particles at one-loop order, it is not necessary to deal with the ghost particles because they are not present in such process up to that order. The propagators of the relevant fields of the model can be cast as
\begin{eqnarray}
\langle T~h^{\alpha\beta}(p) h^{\mu\nu}(-p)\rangle&=&D^{\alpha\beta\mu\nu}(p)=\frac{i}{p^2-2\Lambda}P^{\alpha\beta\mu\nu},\label{eq05}\\
\langle T~\phi(p) \phi(-p)\rangle &=&\Delta(p)=\frac{i}{p^2}.\label{eq05a}
\end{eqnarray}

The main effect of the presence of a cosmological constant in the weak field approximation of the Einstein's gravity is the generation of a massive pole to the graviton propagator~\cite{Hamber:2007fk}, as we see from Eq.(\ref{eq05}). Moreover, in our approach there is a linear term of type $\Lambda h$ in the expanded Lagrangian which can be interpreted as a source for the gravitational field~\cite{Hamber:2007fk}.

Now, let us compute the quantum gravity corrections to the scattering amplitude between two scalar particles in self-interaction. We will use dimensional regularization~\cite{'tHooft:1972fi} to evaluate the integrals and minimal subtraction scheme to renormalize the amplitude.

\section{Amplitude scattering between scalar particles}\label{sec2}

\subsection{Tree level amplitude}

\begin{figure}[h] \includegraphics[width=14cm]{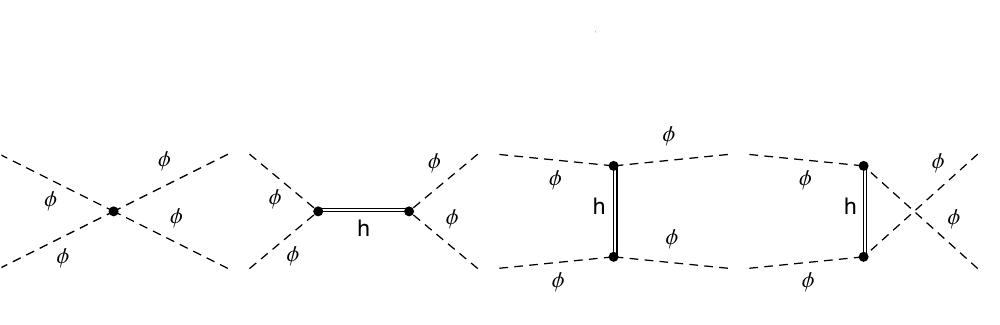}  \caption{Tree level Feynman diagrams which contribute to the scattering amplitude of neutral scalar particles. Dashed and double continuous lines represent the scalar particle and graviton propagators, respectively.} \label{fig1} \end{figure}

The tree level Feynman diagrams of the scattering amplitude between two scalar particles, $\phi+\phi\rightarrow\phi+\phi$, are drawn in Figure \ref{fig1}.  The corresponding amplitude is given by
\begin{eqnarray}
\mathcal{M}_{tree}&=&-\lambda-\frac{\kappa ^2 \left(s~\text{p}_3^2+s ~\text{p}_4^2+\text{p}_3^2 \text{p}_4^2+s ~\text{p}_2^2+\text{p}_1^2 \left(\text{p}_2^2+s\right)+s u+u^2\right)}{s-2 \Lambda}\nonumber\\
&&-\frac{\kappa ^2 \left(\text{p}_1^2 \left(\text{p}_3^2+t\right)+\text{p}_2^2 \left(\text{p}_4^2+t\right)+t~\text{p}_3^2+t~\text{p}_4^2+s^2+s t\right)}{t-2 \Lambda }\\
&&-\frac{\kappa ^2 \left(\text{p}_2^2 \left(\text{p}_3^2+u\right)+\text{p}_1^2 \left(\text{p}_4^2+u\right)+u~\text{p}_3^2+u~\text{p}_4^2+t^2+t u\right)}{u-2 \Lambda}.
\end{eqnarray}

In fact, here we should consider the contribution from the higher order operator $\lambda_1\phi^2\partial^{\mu}\phi\partial_{\mu}\phi$ which is proportional to $\lambda_1(\text{p}_1^2+\text{p}_2^2+\text{p}_3^2+\text{p}_4^2)$. But, to avoid unnecessary complications, we will consider that the coupling $\lambda_1$ is entirely generated by quantum corrections, and without loss of generality we will take $\lambda_1=0$ at tree level. Moreover, we are computing the off-shell scattering amplitude, since the flat space-time is not solution of the Einstein's equations in the presence of a cosmological constant. A similar approach was done in earlier papers~\cite{Toms:2008dq,Pietrykowski:2012nc}. 

\subsection{One-loop corrections}

Let us compute the divergent part of the one-loop corrections to the scattering amplitude $\phi+\phi\rightarrow\phi+\phi$. To do this we use dimensional regularization (which is known to preserve gauge invariance) to evaluate the integrals over the loop momentum and minimal subtraction scheme (MS) of renormalization. Therefore, we have to redefine the coupling constants as 
\begin{eqnarray}
\lambda&=&\mu^\epsilon Z_\lambda\lambda_r=\mu^{\epsilon}(\lambda_r+\delta_\lambda),\nonumber\\
\lambda_1&=&\mu^\epsilon Z_{\lambda_1}{\lambda_1}_r=\mu^{\epsilon}({\lambda_1}_r+\delta_{\lambda_1}), 
\end{eqnarray}

\noindent where $\delta_\lambda$ and $\delta_{\lambda_1}$ are the counterterms and the index $r$ means renormalized quantity, that we will omit hereafter.   
 
The divergent part of the one-loop contribution to the scattering amplitude $\phi+\phi\rightarrow\phi+\phi$, Figure \ref{fig2}, is given by 
\begin{eqnarray}\label{dimred1}
\mathcal{M}_{1loop}&=&\frac{3\lambda^2}{16\pi^2\epsilon}+
\frac{\kappa ^2 \lambda  \left(15 \text{p}_3^2+15 \text{p}_4^2+14 \text{p}_1^2+14 \text{p}_2^2-5 \text{m}_g^2+s\right)}{32 \pi ^2 \epsilon }\nonumber\\
&&+\frac{\kappa ^2 \lambda}{64 \pi ^2 \epsilon  (t-2 \Lambda )}  \Big[\text{p}_1^2 \left(7 \text{p}_3^2+4 \text{p}_4^2+3 \text{p}_2^2-s+3 t\right) +3 \text{p}_2^2 \left(\text{p}_3^2+t\right)\nonumber\\
&&+s \text{p}_3^2+2 t \text{p}_3^2+2 t \text{p}_4^2+(\text{p}_3^2)^2+(\text{p}_4^2)^2+6 \text{p}_3^2 \text{p}_4^2-(\text{p}_2^2)^2+3 s t+4 t^2\Big]\nonumber\\
&&-\frac{\kappa ^2 \lambda}{64 \pi ^2 \epsilon  (u-2 \Lambda )} \Big[\text{p}_2^2 \left(\text{p}_3^2-2 \text{p}_4^2+t-2 u\right)-3 \text{p}_1^2 \left(\text{p}_3^2+2 \text{p}_4^2+\text{p}_2^2+u\right)\nonumber\\
&&+t \text{p}_3^2+2 t \text{p}_4^2-3 u \text{p}_3^2-2 u \text{p}_4^2-4 \text{p}_3^2 \text{p}_4^2+(\text{p}_2^2)^2+t u-3 u^2\Big]\nonumber\\
&&+\frac{\kappa ^2 \lambda}{32 \pi ^2 \epsilon  (s-2 \Lambda )}  \Big[\text{p}_2^2 \left(2 \text{p}_3^2+2 \text{p}_4^2+3 s\right)+\text{p}_1^2 \left(2 \text{p}_3^2+2 \text{p}_4^2+4 \text{p}_2^2+3 s\right)\nonumber\\
&&+s \left(2 \text{p}_3^2+2 \text{p}_4^2+s\right)\Big]+\mathrm{(finite~terms)},
\end{eqnarray}

\noindent where $\epsilon=(4-D)$ ($D$ is the space-time dimension) and $s=(\text{p}_1+\text{p}_2)^2$, $t=(\text{p}_1-\text{p}_3)^2$ and $u=(\text{p}_1-\text{p}_4)^2$ are the Mandelstam variables, with $(\text{p}_1,\text{p}_2)$ and $(\text{p}_3,\text{p}_4)$ being the incoming and outgoing momenta, respectively. Some details of the amplitude construction and evaluation of the integrals over loop momentum are given in the Appendix. 

As we mentioned above, we considered that the coupling $\lambda_1$ is entirely generated by quantum corrections, so $\lambda_1=0$ at tree level. Therefore, the vertex containing $\lambda_1$ does not appear at one-loop process, but we will need the counterterm $\delta_{\lambda_1}$ to renormalize the amplitude properly~\cite{Anber:2010uj}, separating what renormalizes $\lambda$ and what renormalizes the relevant higher order operator, i.e., $\lambda_1$. We choose to keep $\kappa$ fixed, then it will not receive any correction. This choice is based on the arguments given by the authors of Ref.\cite{Anber:2011ut}, which have shown that the running of the gravitational coupling constant should be process dependent.   

Let us subtract the poles from the total amplitude $\mathcal{M}=\mathcal{M}_{tree}+\mathcal{M}_{1loop}$ at the Euclidean momenta $\text{p}_1^2=\text{p}_2^2=\text{p}_3^2=\text{p}_4^2=-M^2$ and $s=t=u=-\frac{4M^2}{3}$. Following a similar suggestion presented in Ref.\cite{Donoghue:2012zc}, let us choose the renormalization conditions as
\begin{eqnarray}
\left(\mathcal{M}\right)\Big{|}_{sp}&=&\left(\mathcal{M}_{tree}\right)\Big{|}_{sp},\label{cond01}\\
\left(\frac{d\mathcal{M}}{dM^2}\right)\Big{|}_{sp}&=&\left(\frac{d\mathcal{M}_{tree}}{dM^2}\right)\Big{|}_{sp},\label{cond02}
\end{eqnarray}
\noindent where $sp$ stands for the quantity between parentheses evaluated at subtraction point. Eqs. (\ref{cond01}) and (\ref{cond02}) are the conditions for the renormalization of the couplings $\lambda$ and $\lambda_1$, respectively.

The renormalization of the higher order operator comes with a factor of $M^2$ when evaluated at $sp$, then its counterterm appearing in the tree level amplitude is $\delta{\lambda_1}M^2$. Having this statement in mind, the renormalization conditions above are given by
\begin{eqnarray}
&&\left(\mathcal{M}\right)\Big{|}_{sp}=\left(\mathcal{M}_{tree}\right)\Big{|}_{sp},\nonumber\\
&&-\delta \lambda -\lambda +\text{$\delta \lambda $1} M^2+\frac{\kappa ^2 M^2}{2}-\frac{17 \kappa ^2 \lambda  M^2}{24 \pi ^2 \epsilon }-\frac{3 \kappa ^2 \Lambda}{4}-\frac{11 \kappa ^2 \lambda  \Lambda}{32 \pi ^2 \epsilon }+\frac{3 \lambda ^2}{16 \pi ^2 \epsilon }\nonumber\\
&&=-\lambda +\frac{\kappa ^2 M^2}{2}-\frac{3 \kappa ^2 \Lambda}{4},\label{cond01a}
\end{eqnarray}

\noindent and
\begin{eqnarray}
&&\left(\frac{d\mathcal{M}}{dM^2}\right)\Big{|}_{sp}=\left(\frac{d\mathcal{M}_{tree}}{dM^2}\right)\Big{|}_{sp},\nonumber\\
&&\text{$\delta \lambda $1}+\frac{\kappa ^2}{2}-\frac{17 \kappa ^2 \lambda }{24 \pi ^2 \epsilon }=\frac{\kappa ^2}{2}.
\label{cond02a}
\end{eqnarray}

From the equations above, we obtain the counterterms
\begin{eqnarray}
\delta_\lambda&=&\frac{3 \lambda ^2}{16 \pi ^2 \epsilon }-\frac{11 \kappa ^2 \lambda}{32 \pi ^2 \epsilon}\Lambda,\nonumber\\
\delta_{\lambda_1}&=&\frac{17 \kappa ^2 \lambda }{24 \pi ^2 \epsilon}.
\end{eqnarray}

Through the knowledge of the counterterms and reminding that the relation between the bare and renormalized coupling is $\lambda_0=\mu^{\epsilon} Z_\lambda \lambda=\mu^{\epsilon}(\lambda+\delta_\lambda)$, we can evaluate the $\beta$-functions of the couplings $\lambda$ and $\lambda_1$. Considering $\kappa$ and $\Lambda$ fixed, and since the bare couplings do not depend on $\mu$, we obtain
\begin{eqnarray}
\beta(\lambda)&=&\frac{3 \lambda ^2}{16 \pi^2}-\frac{11 \kappa^2 \lambda}{32\pi^2}\Lambda,\label{dimred5a}\\
\beta(\lambda_1)&=&\frac{17 \kappa^2 \lambda }{24 \pi^2}.\label{dimred5b}
\end{eqnarray}

The beta function given in Eq.(\ref{dimred5a}) suggests that gravitational (in the presence of a cosmological constant) can change the running behavior of $\lambda$, protecting it from quantum triviality. Of course, this can not be considered a solution to the problem involving the Higgs sector of the SM, but we believe it seems like a good direction to search for answers. This effect due to cosmological constant is similar to that obtained for the electric charge in Ref.\cite{Toms:2011zza} through effective action techniques. If we put the cosmological constant to vanish this effect is destroyed and gravitational corrections only renormalize a higher order operator. 

\section{Concluding Remarks}\label{conc}

In summary, we have evaluated the one-loop gravitational corrections to the scattering of scalar particles up to $\mathcal{O}(\kappa^2)$. For a fixed gravitational coupling constant $\kappa$ and the cosmological constant, we have shown that gravitational corrections to the beta function of $\lambda$ is a negative term proportional to $\Lambda$, therefore driving $\lambda$ to run in the direction of asymptotic freedom. Considering the magnitude of the observed $\Lambda$ ($\sim 10^{-47}$GeV$^4$), such effect has no practical phenomenological consequence, but it indicates that a complete theory of quantum gravity should preserve field theories from quantum triviality. In addition, we have not considered massive scalar particles. When massive particles are considered, a Feynman diagrammatic approach presented in~\cite{Rodigast:2009zj} indicates that gravitational corrections to the running of coupling constant $\lambda$ also present a running behavior of asymptotic freedom. This type of effect, that is proportional to the mass squared, is more severe than that from cosmological constant. We think that the computation of the scattering amplitude for massive particles should corroborate such result.
\vspace{.5cm}

{\bf Acknowledgments.}
This work was partially supported by Funda\c{c}\~ao de Amparo \`a Pesquisa do Estado de S\~ao Paulo (FAPESP), Conselho Nacional de Desenvolvimento Cient\'{\i}fico e Tecnol\'{o}gico (CNPq) and Funda\c{c}\~{a}o de Apoio \`{a} Pesquisa do Rio Grande do Norte (FAPERN). The author would like to thank L. Ibiapina Bevilaqua for the careful reading and A. J. da Silva for the useful discussions.


\section*{Appendix: One-loop scattering amplitudes} 

\begin{figure}[h] \includegraphics[width=14cm]{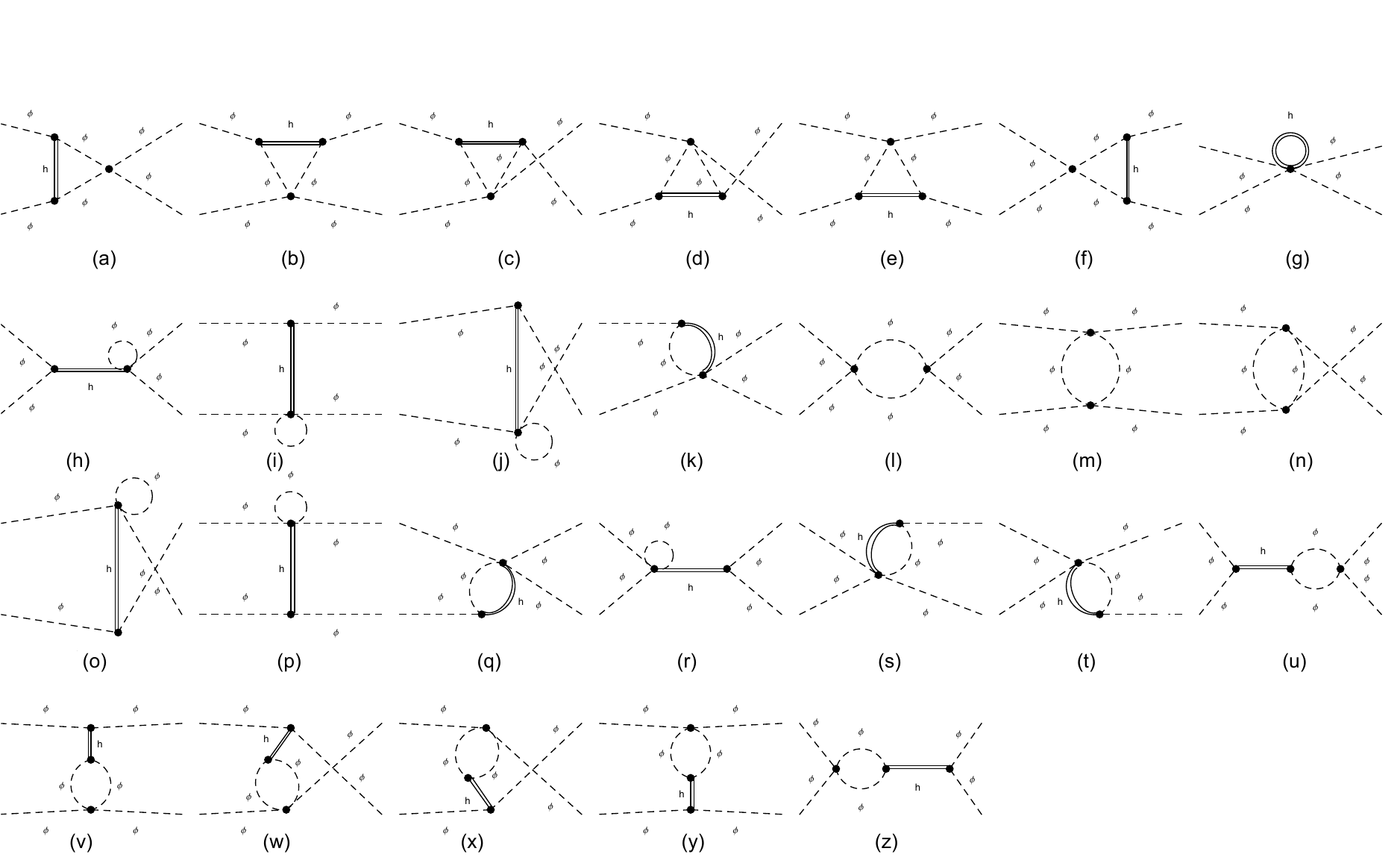}  \caption{One-loop contributions to the scattering amplitude of neutral scalar particles up to $\mathcal{O}(\kappa^2)$.} \label{fig2} \end{figure}

Let us give some details about the calculation of the one-loop corrections to the scattering amplitude between two scalar particles, Figure \ref{fig2}. To evaluate the integrals over internal momentum $\text{q}$, we have expanded the integrand for small values of external momenta $\text{p}_i$ and used dimensional regularization to calculate them. In fact, we consider massive scalar particles to regulate some potentially IR divergence and take the limit $\text{m}\rightarrow0$ in the end of calculation. 

At one-loop are twenty-six Feynman diagrams that contribute to the process, we limited ourselves to evaluate the quantum gravitational correction up to order of $\kappa^2$, corresponding to processes with one graviton exchanging. To build and manipulate the amplitude, we used the Mathematica$^{\copyright}$ packages \emph{FeynArts}~\cite{feynarts} and \emph{FeynCalc}~\cite{feyncalc}.

The corresponding expressions to the one-loop diagrams Figure \ref{fig2} are given by 
\begin{eqnarray}
\mathcal{M}_{(a)}&=&\int
\frac{d^4\text{q}}{2 \pi ^4}~3 i \kappa ^2 \lambda \frac{1}{\text{q}^2-\text{m}_g^2}\frac{1}{(\text{p}_2+\text{q})^2-\text{m}^2}\frac{1}{(-\text{p}_3-\text{p}_4+\text{p}_2+\text{q})^2-\text{m}^2}\nonumber\\
&& \Big[\text{p}_1\cdot \text{p}_4 \text{p}_2\cdot \text{q}+\text{p}_1\cdot \text{p}_3 \left(\text{p}_2\cdot \text{q}+\text{p}_2^2\right)-\text{p}_2\cdot \text{p}_3 \text{p}_1\cdot \text{q}-\text{p}_2\cdot \text{p}_4 \text{p}_1\cdot \text{q}\nonumber\\
&&+\text{p}_1\cdot \text{p}_2 \left(-2 \text{p}_2\cdot \text{p}_3-2 \text{p}_2\cdot \text{p}_4-\text{q}\cdot \text{p}_3-\text{q}\cdot \text{p}_4+2 \text{p}_2\cdot \text{q}+\text{p}_2^2+\text{q}^2\right)+\text{p}_2^2 \text{p}_1\cdot \text{p}_4\Big]\nonumber\\
&=&-\frac{3 \kappa ^2 \lambda  \text{p}_1\cdot \text{p}_2}{\pi ^2 \epsilon}+(\text{finite});
\end{eqnarray}
\begin{eqnarray}
\mathcal{M}_{(b)}&=&-\int
\frac{d^4\text{q}}{2 \pi ^4}~i \kappa ^2 \lambda\left(\frac{1}{\text{q}^2-\text{m}_g^2}\frac{1}{(\text{q}-\text{p}_3)^2-\text{m}^2}\frac{1}{(-\text{p}_3-\text{p}_4+\text{p}_2+\text{q})^2-\text{m}^2}\right)\nonumber\\
&& \Big[-(\text{p}_1\cdot \text{p}_2-\text{p}_1\cdot \text{p}_4) \left(\text{q}\cdot \text{p}_3-\text{p}_3^2\right)+(\text{p}_2\cdot \text{p}_3-\text{p}_3\cdot \text{p}_4) \text{p}_1\cdot \text{q}\nonumber\\
&&+\text{p}_1\cdot \text{p}_3 \left(-2 \text{p}_2\cdot \text{p}_3-2 \text{q}\cdot \text{p}_3-\text{q}\cdot \text{p}_4+\text{p}_3^2+2 \text{p}_3\cdot \text{p}_4+\text{p}_2\cdot \text{q}+\text{q}^2\right)\Big]\nonumber\\
&=&\frac{3 \kappa ^2 \lambda  \text{p}_1\cdot \text{p}_3}{\pi ^2 \epsilon}+(\text{finite});
\end{eqnarray}
\begin{eqnarray}
\mathcal{M}_{(c)}&=&-\int
\frac{d^4\text{q}}{2 \pi ^4}~3 i \kappa ^2 \lambda \left(\frac{1}{\text{q}^2-\text{m}_g^2}\frac{1}{(\text{q}-\text{p}_4)^2-\text{m}^2}\frac{1}{(-\text{p}_3-\text{p}_4+\text{p}_2+\text{q})^2-\text{m}^2}\right)\nonumber\\
&&  \Big[(\text{p}_2\cdot \text{p}_4-\text{p}_3\cdot \text{p}_4) \text{p}_1\cdot \text{q}-(\text{p}_1\cdot \text{p}_2-\text{p}_1\cdot \text{p}_3) \left(\text{q}\cdot \text{p}_4-\text{p}_4^2\right)\nonumber\\
&&+\text{p}_1\cdot \text{p}_4 \left(-2 \text{p}_2\cdot \text{p}_4-\text{q}\cdot \text{p}_3-2 \text{q}\cdot \text{p}_4+\text{p}_4^2+2 \text{p}_3\cdot \text{p}_4+\text{p}_2\cdot \text{q}+\text{q}^2\right)\Big] \nonumber\\
&=&\frac{3 \kappa ^2 \lambda  \text{p}_1\cdot \text{p}_4}{\pi ^2 \epsilon}+(\text{finite});
\end{eqnarray}
\begin{eqnarray}
\mathcal{M}_{(d)}&=&-\int
\frac{d^4\text{q}}{2 \pi ^4}~3 i \kappa ^2 \lambda\left(\frac{1}{\text{q}^2-\text{m}_g^2}\frac{1}{(\text{q}-\text{p}_2)^2-\text{m}^2}\frac{1}{(\text{q}-\text{p}_3)^2-\text{m}^2}\right)\nonumber\\
&&  \Big[\text{p}_2^2 \left(\text{q}\cdot \text{p}_3-\text{p}_3^2+\text{m}^2\right)+\text{p}_2\cdot \text{q} \left(-2 \text{p}_2\cdot \text{p}_3+\text{p}_3^2\right)+\text{q}^2 \text{p}_2\cdot \text{p}_3\nonumber\\
&&-2 \text{p}_2\cdot \text{p}_3 \text{q}\cdot \text{p}_3+2 \text{p}_2\cdot \text{p}_3^2\Big] \nonumber\\
&=&\frac{3 \kappa ^2 \lambda  \text{p}_2\cdot \text{p}_3}{\pi ^2 \epsilon}+(\text{finite});
\end{eqnarray}
\begin{eqnarray}
\mathcal{M}_{(e)}&=&-\int
\frac{d^4\text{q}}{2 \pi ^4}~3 i \kappa ^2 \lambda \left(\frac{1}{\text{q}^2-\text{m}_g^2}\frac{1}{(\text{q}-\text{p}_2)^2-\text{m}^2}\frac{1}{(\text{q}-\text{p}_4)^2-\text{m}^2}\right)\nonumber\\
&& \Big[\text{p}_2\cdot \text{p}_4 \left(2 \text{p}_2\cdot \text{p}_4-2 \text{q}\cdot \text{p}_4-2 \text{p}_2\cdot \text{q}+\text{q}^2\right)+\text{p}_2^2 \left(\text{q}\cdot \text{p}_4-\text{p}_4^2\right)+\text{p}_4^2 \text{p}_2\cdot \text{q}\Big] \nonumber\\
&=&\frac{3 \kappa ^2 \lambda  \text{p}_2\cdot \text{p}_4}{\pi ^2 \epsilon}+(\text{finite});
\end{eqnarray}
\begin{eqnarray}
\mathcal{M}_{(f)}&=&\int
\frac{d^4\text{q}}{2 \pi ^4}~3 i \kappa ^2 \lambda \left(\frac{1}{\text{q}^2-\text{m}_g^2}\frac{1}{(\text{p}_3+\text{q})^2-\text{m}^2}\frac{1}{(\text{q}-\text{p}_4)^2-\text{m}^2}\right) \nonumber\\
&&\Big[\text{p}_3\cdot \text{p}_4 \left(2 \text{q}\cdot \text{p}_3-2 \text{p}_3\cdot \text{p}_4+\text{q}^2\right)-\left(\text{p}_3^2+2 \text{p}_3\cdot \text{p}_4\right) \text{q}\cdot \text{p}_4+\text{p}_4^2 \left(\text{q}\cdot \text{p}_3+\text{p}_3^2\right)\Big] \nonumber\\
&=&-\frac{3 \kappa ^2 \lambda  \text{p}_3\cdot \text{p}_4}{\pi ^2 \epsilon}+(\text{finite});
\end{eqnarray}
\begin{eqnarray}
\mathcal{M}_{(g)}&=&\int
\frac{d^4\text{q}}{16 \pi ^4}\frac{15 i \kappa ^2 \lambda }{(\text{q}^2-\text{m}_g^2)} =-\frac{15 \kappa ^2 \lambda  \text{m}_g^2}{8\pi ^2 \epsilon}+(\text{finite});
\end{eqnarray}
\begin{eqnarray}
\mathcal{M}_{(h)}&=&\int
\frac{d^4\text{q}}{8 \pi ^4}~3 i \kappa ^2 \lambda \frac{\text{p}_1\cdot \text{p}_2}{(\text{p}_3+\text{p}_4)^2-\text{m}_g^2}\left(\frac{1}{\text{q}^2-\text{m}^2}\right)\nonumber\\
&=&-\frac{3 \kappa ^2 \lambda  \text{m}^2}{4 \pi ^2 \epsilon}\frac{\left(\text{p}_1\cdot \text{p}_2\right)}{(\text{p}_3+\text{p}_4)^2-\text{m}_g^2}+(\text{finite})=0~(\text{massless~limit});
\end{eqnarray}
\begin{eqnarray}
\mathcal{M}_{(i)}&=&-\int
\frac{d^4\text{q}}{8 \pi ^4}~3 i \kappa ^2 \lambda \frac{\left(\text{p}_1\cdot \text{p}_3\right)}{(\text{p}_4-\text{p}_2)^2-\text{m}_g^2}\left(\frac{1}{\text{q}^2-\text{m}^2}\right)\nonumber\\
&=&\frac{3 \kappa ^2 \lambda  \text{m}^2}{4 \pi ^2 \epsilon}\frac{\left(\text{p}_1\cdot \text{p}_3\right)}{(\text{p}_4-\text{p}_2)^2-\text{m}_g^2}+(\text{finite})=0~(\text{massless~limit});
\end{eqnarray}
\begin{eqnarray}
\mathcal{M}_{(j)}&=&-\int
\frac{d^4\text{q}}{8 \pi ^4}~3 i \kappa ^2 \lambda \frac{\left(\text{p}_1\cdot \text{p}_4\right)}{(\text{p}_4-\text{p}_2)^2-\text{m}_g^2}\left(\frac{1}{\text{q}^2-\text{m}^2}\right)\nonumber\\
&=&\frac{3 \kappa ^2 \lambda  \text{m}^2}{4 \pi ^2 \epsilon}\frac{\left(\text{p}_1\cdot \text{p}_4\right)}{(\text{p}_4-\text{p}_2)^2-\text{m}_g^2}=0~(\text{massless~limit});
\end{eqnarray}
\begin{eqnarray}
\mathcal{M}_{(k)}&=&-\int
\frac{d^4\text{q}}{4 \pi ^4}~3 i \kappa ^2 \lambda  \text{p}_1\cdot \text{q}\left(\frac{1}{\text{q}^2-\text{m}^2}\frac{1}{(-\text{p}_3-\text{p}_4+\text{p}_2+\text{q})^2-\text{m}_g^2}\right)\nonumber\\
&=&\frac{3 \kappa ^2 \lambda  \left(\text{p}_1\cdot \text{p}_3+\text{p}_1\cdot \text{p}_4\right)}{4 \pi ^2 \epsilon}+(\text{finite});
\end{eqnarray}
\begin{eqnarray}
\mathcal{M}_{(l)}&=&-\int
d^4\text{q}~\frac{9 i \lambda ^2}{2 \pi ^4}\left(\frac{1}{\text{q}^2-\text{m}^2}\frac{1}{(-\text{p}_3-\text{p}_4+\text{q})^2-\text{m}^2}\right)\nonumber\\
&=&\frac{9 \lambda ^2}{\pi ^2 \epsilon}+(\text{finite});
\end{eqnarray}
\begin{eqnarray}
\mathcal{M}_{(m)}&=&-\int
d^4\text{q}~\frac{9 i \lambda ^2}{2 \pi ^4}\left(\frac{1}{\text{q}^2-\text{m}^2}\frac{1}{(-\text{p}_4+\text{p}_2+\text{q})^2-\text{m}^2}\right)\nonumber\\
&=&\frac{9 \lambda ^2}{\pi ^2 \epsilon}+(\text{finite});
\end{eqnarray}
\begin{eqnarray}
\mathcal{M}_{(n)}&=&-\int
d^4\text{q}~\frac{9 i \lambda ^2}{2 \pi ^4}\left(\frac{1}{\text{q}^2-\text{m}^2}\frac{1}{(-\text{p}_3+\text{p}_2+\text{q})^2-\text{m}^2}\right)\nonumber\\
&=&\frac{9 \lambda ^2}{\pi ^2 \epsilon}+(\text{finite});
\end{eqnarray}
\begin{eqnarray}
\mathcal{M}_{(o)}&=&-\int\frac{d^4\text{q}}{8 \pi ^4}3 i \kappa ^2 \lambda
\frac{\text{p}_2\cdot \text{p}_3}{(\text{p}_2-\text{p}_3)^2-\text{m}_g^2}
\left(\frac{1}{\text{q}^2-\text{m}^2}\right) 
\nonumber\\
&=&\frac{3 \kappa ^2 \lambda  \text{m}^2}{4 \pi ^2 \epsilon}\frac{\left(\text{p}_2\cdot \text{p}_3\right)}{(\text{p}_2-\text{p}_3)^2-\text{m}_g^2}=0~(\text{massless~limit});
\end{eqnarray}
\begin{eqnarray}
\mathcal{M}_{(p)}&=&-\int\frac{d^4\text{q}}{4 \pi ^4}3 i \kappa ^2 \lambda
\frac{\left(\text{p}_2\cdot \text{p}_4\right)}{(\text{p}_2-\text{p}_4)^2-\text{m}_g^2}
\left(\frac{1}{\text{q}^2-\text{m}^2}\right) 
\nonumber\\
&=&\frac{3 \kappa ^2 \lambda  \text{m}^2}{4 \pi ^2 \epsilon}\frac{\left(\text{p}_2\cdot \text{p}_4\right)}{(\text{p}_2-\text{p}_4)^2-\text{m}_g^2}=0~(\text{massless~limit});
\end{eqnarray}
\begin{eqnarray}
\mathcal{M}_{(q)}&=&-\int\frac{d^4\text{q}}{4 \pi ^4}3 i \kappa ^2 \lambda
\text{p}_2\cdot \text{q}\left(\frac{1}{\text{q}^2-\text{m}^2}\frac{1}{(\text{q}-\text{p}_2)^2-\text{m}_g^2}\right)
\nonumber\\
&=&\frac{3 \kappa ^2 \lambda  \text{p}_2^2}{4 \pi ^2 \epsilon}+(\text{finite});
\end{eqnarray}
\begin{eqnarray}
\mathcal{M}_{(r)}&=&\int\frac{d^4\text{q}}{8 \pi ^4}~3 i \kappa ^2 \lambda
\frac{\text{p}_3\cdot \text{p}_4}{(\text{p}_3+\text{p}_4)^2-\text{m}_g^2} \left(\frac{1}{\text{q}^2-\text{m}^2}\right)
\nonumber\\
&=&-\frac{3 \kappa ^2 \lambda  \text{m}^2}{4 \pi ^2 \epsilon}\frac{\text{p}_3\cdot \text{p}_4}{(\text{p}_3+\text{p}_4)^2-\text{m}_g^2}=0~(\text{massless~limit});
\end{eqnarray}
\begin{eqnarray}
\mathcal{M}_{(s)}&=&\int\frac{d^4\text{q}}{4 \pi ^4}~3 i \kappa ^2 \lambda
\text{q}\cdot \text{p}_3 \left(\frac{1}{\text{q}^2-\text{m}^2}\frac{1}{(\text{p}_3+\text{q})^2-\text{m}_g^2}\right)\nonumber\\
&=&\frac{3 \kappa ^2 \lambda  \text{p}_3^2}{4 \pi ^2 \epsilon}+(\text{finite});
\end{eqnarray}
\begin{eqnarray}
\mathcal{M}_{(t)}&=&\int\frac{d^4\text{q}}{4 \pi ^4}~3 i \kappa ^2 \lambda
\text{q}\cdot \text{p}_4 \left(\frac{1}{\text{q}^2-\text{m}^2}\frac{1}{(\text{p}_4+\text{q})^2-\text{m}_g^2}\right)\nonumber\\
&=&\frac{3 \kappa ^2 \lambda  \text{p}_4^2}{4 \pi ^2 \epsilon}+(\text{finite});
\end{eqnarray}
\begin{eqnarray}
\mathcal{M}_{(u)}&=&-\int\frac{d^4\text{q}}{4 \pi ^4}~3 i \kappa ^2 \lambda
\frac{1}{(\text{p}_3+\text{p}_4)^2-\text{m}_g^2}
\Big[(\text{p}_1\cdot \text{p}_3+\text{p}_1\cdot \text{p}_4) \text{p}_2\cdot \text{q}\nonumber\\
&&+\text{p}_1\cdot \text{q} (\text{p}_2\cdot \text{p}_3+\text{p}_2\cdot \text{p}_4-2 \text{p}_2\cdot \text{q})+\text{p}_1\cdot \text{p}_2 \left(-\text{q}\cdot \text{p}_3-\text{q}\cdot \text{p}_4+\text{q}^2\right)\Big]
\nonumber\\
&&\left(\frac{1}{\text{q}^2-\text{m}^2}\frac{1}{(-\text{p}_3-\text{p}_4+\text{q})^2-\text{m}^2}\right)\nonumber\\
&=&-\frac{3 \kappa ^2 \lambda}{4 \pi ^2 \epsilon}\frac{\left(\text{p}_3^2+\text{p}_4^2+2 \text{p}_3\cdot \text{p}_4\right) \left(\text{p}_1\cdot \text{p}_2\right)-2 (\text{p}_1\cdot \text{p}_3+\text{p}_1\cdot \text{p}_4) (\text{p}_2\cdot \text{p}_3+\text{p}_2\cdot \text{p}_4)}{(\text{p}_3+\text{p}_4)^2-\text{m}_g^2}\nonumber\\
&&+(\text{finite});
\end{eqnarray}
\begin{eqnarray}
\mathcal{M}_{(v)}&=&\int\frac{d^4\text{q}}{4 \pi ^4}~3 i \kappa ^2 \lambda
\frac{1}{(\text{p}_4-\text{p}_2)^2-\text{m}_g^2}
\Big[(\text{p}_1\cdot \text{p}_4-\text{p}_1\cdot \text{p}_2) \text{q}\cdot \text{p}_3\nonumber\\
&&+\text{p}_1\cdot \text{p}_3 \left(-\text{q}\cdot \text{p}_4+\text{p}_2\cdot \text{q}+\text{q}^2\right)+\text{p}_1\cdot \text{q} (-(\text{p}_2\cdot \text{p}_3)-2 \text{q}\cdot \text{p}_3+\text{p}_3\cdot \text{p}_4)\Big]
\nonumber\\
&&\left(\frac{1}{\text{q}^2-\text{m}^2}\frac{1}{(-\text{p}_4+\text{p}_2+\text{q})^2-\text{m}^2}\right)\nonumber\\
&=&-\frac{3 \kappa ^2 \lambda}{4 \pi ^2 \epsilon}\frac{1}{(\text{p}_4-\text{p}_2)^2-\text{m}_g^2}\Big[\left(\text{p}_3\cdot \text{p}_4\right) \text{p}_1\cdot \text{p}_3+\text{p}_1\cdot \text{p}_2 \text{p}_2\cdot \text{p}_3\nonumber\\
&&-2 \text{p}_1\cdot \text{p}_4 \text{p}_2\cdot \text{p}_3-\text{p}_3\cdot \text{p}_4 \text{p}_1\cdot \text{p}_2+2 \text{p}_3\cdot \text{p}_4 \text{p}_1\cdot \text{p}_4\Big]+(\text{finite});
\end{eqnarray}
\begin{eqnarray}
\mathcal{M}_{(w)}&=&\int\frac{d^4\text{q}}{4 \pi ^4}~3 i \kappa ^2 \lambda
\frac{1}{(\text{p}_3-\text{p}_2)^2-\text{m}_g^2}
\Big[\text{p}_1\cdot \text{p}_4 \left(-\text{q}\cdot \text{p}_3+\text{p}_2\cdot \text{q}+\text{q}^2\right)\nonumber\\
&&+(\text{p}_1\cdot \text{p}_3-\text{p}_1\cdot \text{p}_2) \text{q}\cdot \text{p}_4+\text{p}_1\cdot \text{q} (-(\text{p}_2\cdot \text{p}_4)-2 \text{q}\cdot \text{p}_4+\text{p}_3\cdot \text{p}_4)\Big]
\nonumber\\
&&\left(\frac{1}{\text{q}^2-\text{m}^2}\frac{1}{(-\text{p}_3+\text{p}_2+\text{q})^2-\text{m}^2}\right)\nonumber\\
&=&-\frac{3 \kappa ^2 \lambda}{4 \pi ^2 \epsilon}\frac{\text{p}_1\cdot \text{p}_2 \text{p}_2\cdot \text{p}_4-2 \text{p}_1\cdot \text{p}_3 \text{p}_2\cdot \text{p}_4-\text{p}_3\cdot \text{p}_4 \text{p}_1\cdot \text{p}_2+2 \text{p}_3\cdot \text{p}_4 \text{p}_1\cdot \text{p}_3}{(\text{p}_3-\text{p}_2)^2-\text{m}_g^2}\nonumber\\
&&+(\text{finite});
\end{eqnarray}
\begin{eqnarray}
\mathcal{M}_{(x)}&=&\int\frac{d^4\text{q}}{4 \pi ^4}~3 i \kappa ^2 \lambda
\frac{\text{q}^2 \text{p}_2\cdot \text{p}_3+\text{p}_2^2 \text{q}\cdot \text{p}_3-\left(2 \text{q}\cdot \text{p}_3+\text{p}_3^2\right) \text{p}_2\cdot \text{q}}{(\text{p}_2-\text{p}_3)^2-\text{m}_g^2}
\nonumber\\
&&\left(\frac{1}{\text{q}^2-\text{m}^2}\frac{1}{(\text{p}_3-\text{p}_2+\text{q})^2-\text{m}^2}\right)\nonumber\\
&=&-\frac{3 \kappa ^2 \lambda}{4 \pi ^2 \epsilon}\frac{2 \text{p}_3^2 \text{p}_2\cdot \text{p}_3+\text{p}_2^2 \left(2 \text{p}_2\cdot \text{p}_3\right)-2 \text{p}_2\cdot \text{p}_3^2}{(\text{p}_2-\text{p}_3)^2-\text{m}_g^2}+(\text{finite});
\end{eqnarray}
\begin{eqnarray}
\mathcal{M}_{(y)}&=&\int\frac{d^4\text{q}}{4 \pi ^4}~3 i \kappa ^2 \lambda
\frac{\text{q}^2 \text{p}_2\cdot \text{p}_4+\text{p}_2^2 \text{q}\cdot \text{p}_4-\left(2 \text{q}\cdot \text{p}_4+\text{p}_4^2\right) \text{p}_2\cdot \text{q}}{(\text{p}_2-\text{p}_4)^2-\text{m}_g^2}
\nonumber\\
&&\left(\frac{1}{\text{q}^2-\text{m}^2}\frac{1}{(\text{p}_4-\text{p}_2+\text{q})^2-\text{m}^2}\right)\nonumber\\
&=&\frac{3 \kappa ^2 \lambda}{4 \pi ^2 \epsilon}\frac{2 \text{p}_2\cdot \text{p}_4^2-\left(2 \text{p}_4^2+\text{p}_3\cdot \text{p}_4\right) \text{p}_2\cdot \text{p}_4-\text{p}_2^2 \left(2 \text{p}_2\cdot \text{p}_4\right)}{(\text{p}_2-\text{p}_4)^2-\text{m}_g^2}+(\text{finite});
\end{eqnarray}
\begin{eqnarray}
\mathcal{M}_{(z)}&=&\int\frac{d^4\text{q}}{4 \pi ^4}~3 i \kappa ^2 \lambda
\frac{\text{p}_3^2 \text{q}\cdot \text{p}_4-\text{q}^2 \text{p}_3\cdot \text{p}_4+\text{q}\cdot \text{p}_3 \left(2 \text{q}\cdot \text{p}_4+\text{p}_4^2\right)}{(\text{p}_3+\text{p}_4)^2-\text{m}_g^2} 
\nonumber\\
&&\left(\frac{1}{\text{q}^2-\text{m}^2}\frac{1}{(\text{p}_3+\text{p}_4+\text{q})^2-\text{m}^2}\right)\nonumber\\
&=&\frac{3 \kappa ^2 \lambda}{4 \pi ^2 \epsilon}\frac{2 \text{p}_3\cdot \text{p}_4 \text{p}_4^2+\text{p}_3^2 \left(2 \text{p}_3\cdot \text{p}_4\right)+2 \text{p}_3\cdot \text{p}_4^2}{(\text{p}_3+\text{p}_4)^2-\text{m}_g^2}+(\text{finite});
\end{eqnarray}
\noindent where $\epsilon=(4-D)$ ($D$ is the dimension of space-time), $\text{m}_g^2=2\Lambda$ and in the last step we have taken the limit $\text{m}\rightarrow0$.

The result of the sum of all one-loop contributions is given in Eq.(\ref{dimred1}).

\end{document}